\documentstyle[12pt]{article}

\begin{document}

\setcounter{page}{1}

\pagestyle{plain} \vspace{1cm}
\begin{center}
\Large{\bf A Braneworld Dark Energy Model with Induced Gravity and the Gauss-Bonnet Effect}\\
\small \vspace{1cm} {\bf Kourosh Nozari} \quad and \quad {\bf Narges Rashidi }\\
\vspace{0.5cm} {\it Department of Physics, Faculty of Basic
Sciences,\\
University of Mazandaran,\\
P. O. Box 47416-95447, Babolsar, IRAN\\
knozari@umz.ac.ir}

\end{center}
\vspace{1.5cm}
\begin{abstract}
We construct a holographic dark energy model with a non-minimally
coupled scalar field on the brane where Gauss-Bonnet and Induced
Gravity effects are taken into account. This model provides a wide
parameter space with several interesting cosmological implications.
Especially, the equation of state parameter of the model crosses the
phantom divide line and it is possible to realize bouncing solutions
in this setup.
\end{abstract}
\newpage
\section{Introduction}
Recent observational data from CMB temperature fluctuations
spectrum, Supernova type Ia redshift-distance surveys and other data
sources, have shown that the universe is currently in a positively
accelerated phase of expansion and its spatial geometry is nearly
flat\, [1]. Reconciling to Einstein's field equations,
$G_{\mu\nu}=8\pi G T_{\mu\nu}$, we see that to explain these
important achievements, we need either to modify the matter part of
the theory by incorporating an additional cosmological component as
dark energy ( that is, $G_{\mu\nu}=8\pi G
[T^{(M)}_{\mu\nu}+T^{(Dark)}_{\mu\nu}]$\, where $T^{(M)}_{\mu\nu}$
and $T^{(Dark)}_{\mu\nu}$ are energy-momentum tensor of ordinary
matter and dark energy component respectively) or modify the
geometric part of the field equations as the manifestation of the
"dark geometry"  ( that is, $G_{\mu\nu}+G^{(Dark)}_{\mu\nu}=8\pi G
T^{(M)}_{\mu\nu}$\,)[2]. Multi-component dark energy with at least
one non-canonical phantom field is a possible candidate of the first
alternative. This viewpoint has been studied extensively in
literature [3]. Extension of the general relativity to more general
theories on cosmological scales is the basis of the second
alternative ( see for instance [4,5] ). The DGP (
Dvali-Gabadadze-Porrati) braneworld scenario as an infra-red (IR)
modification of the general relativity, explains accelerated
expansion of the universe in its self-accelerating branch via
leakage of gravity to extra dimension [6]. In this model, the
late-time acceleration of the universe is driven by the
manifestation of the excruciatingly slow leakage of gravity off our
four-dimensional braneworld into an extra dimension. \\

There are some datasets (such as the Gold dataset) that show a mild
trend for crossing of the cosmological constat line by the equation
of state (EoS) parameter of dark component in the first alternative.
The equation of state parameter in these scenarios crosses the
phantom divide ( $\omega=\frac{p}{\rho}=-1$) line at recent
redshifts and current accelerated expansion requires
$\omega<-\frac{1}{3}$. The current best fit value of the equation of
state parameter, using WMAP five year data combined with
measurements of type Ia supernovae and Baryon Acoustic Oscillations
( BAO ) in the galaxy distribution, is given by $-0.11 < 1+\omega <
0.14$ ( with $95$ percent CL uncertainties) ( see the paper by
Komatsu {\it et al.} in Ref. [1]). It is accepted that crossing of
the phantom divide line occurs at recent epoch with $z\sim 0.25$
[3].

In the second alternative and within self-accelerating branch of the
DGP scenario, the EoS parameter never crosses the $\omega(z)=-1$
line, and universe eventually turns out to be de Sitter phase.
However, in this setup if we use a single scalar field (ordinary or
phantom) on the brane ( in both branches of the scenario), we can
show that EoS parameter of dark energy can cross the phantom divide
line [7,8]. In a braneworld setup with induced gravity embedded in a
bulk with arbitrary matter content, the transition from a period of
domination of the matter energy density by non-relativistic brane
matter to domination by the generalized dark radiation provides a
crossing of the phantom divide line [9]. Quintessential scheme can
also be achieved in a geometrical way in higher order theories of
gravity [10]( see also [4]). Phantom-like behavior without phantom
matter [8] and also phantom-like behavior in a brane-world setup
with induced gravity and curvature effects have been reported [11].

In a braneworld scenario, the radiative corrections in the bulk lead
to higher curvature terms on the brane. At high energies, the
Einstein-Hilbert action will acquire quantum corrections. The
Gauss-Bonnet (GB) combination arises as the leading bulk correction
in the case of the heterotic string theory [12]. This term leads to
second-order gravitational field equations linear in the second
derivatives in the bulk metric which is ghost free [13-15], the
property of curvature invariant of the Gauss-Bonnet term. Inclusion
of the Gauss-Bonnet term in the action results in a variety of novel
phenomena which certainly affects the cosmological consequences of
these generalized braneworld setup, although these corrections are
smaller than the usual Einstein-Hilbert terms [16-21]. We note that
in a DGP+GB scenario which we call it GBIG, while induced gravity is
a manifestation of the IR limit of the model, the Gauss-Bonnet term
is essentially related to the UV limit of the scenario.

With these preliminaries, the purpose of this paper is to construct
a holographic dark energy model in a DGP setup with induced gravity
and Gauss-Bonnet effect. We consider a scalar field non-minimally
coupled to induced gravity on the brane as a dark energy component
in the presence of radiative corrections. We study a general case in
which a DGP brane with non-vanishing tension is embedded in an AdS
bulk and a scalar field non-minimally coupled to induced gravity is
presented on the brane. We study dynamics of the equation of state
parameter of holographic dark energy component in this setup. Due to
a wide parameter space, this model accounts for crossing of the
phantom divide line and it is possible to realize bouncing solutions
in this framework ( for a review of bouncing solutions see [22]). A
crucial point should be stressed here: the Gauss-Bonnent braneworld
scenario with induced gravity essentially dose not need to introduce
any scalar field to realize phantom divide line crossing. In other
words, the combination of the effect of the Gauss-Bonnent term in
the bulk and induced gravity term on the brane behaves itself as a
dark energy on the brane [14]. However, incorporation of a scalar
field non-minimally coupled to induced gravity on the brane leads to
several interesting features such as shift of the initial conditions
for finite big bang scenario and possibility to realize bouncing
solutions [21]. As we will show, inclusion of this non-minimally
coupled scalar field on the brane brings several new consequences in
the spirit of the holographic dark energy model. So, due to apparent
gap in the study of the holographic dark energy with induced gravity
and Gauss-Bonnet term and also to have a complete picture of these
types of scenarios, this paper considers a scalar field which is
non-minimally coupled to the induced gravity in the presence of the
Gauss-Bonnet term in the bulk. Finally, we study some especial cases
by adopting suitable ansatz and also we address the issue of ghost
instabilities in this setup.

\section{A Non-Minimally Coupled Scalar field on the DGP Brane}
The action of the DGP scenario in the presence of a non-minimally
coupled scalar field on the brane can be written as follows [21]
$$S=\frac{1}{2\kappa_{5}^{2}}\int d^{5}x\sqrt{-g^{(5)}}\Big[
R^{(5)}-2\Lambda_{5}\Big]$$
\begin{equation}
+\Bigg[\frac{r}{2\kappa_{5}^{2}}\int
d^{4}x\sqrt{-g}\bigg(\alpha(\phi) R -2\kappa_{4}^{2} g^{\mu\nu}
\nabla_{\mu}\phi\nabla_{\nu}\phi -4\kappa_{4}^{2}V(\phi) -4
\kappa_{4}^{2}\lambda\bigg)\Bigg]_{y=0},
\end{equation}
where we have included a general non-minimal coupling $\alpha(\phi)$
in the brane part of the action. $y$ is coordinate of the fifth
dimension and we assume the brane is located at $y=0$.\,
$g^{(5)}_{AB}$ is five dimensional bulk metric with Ricci scalar
${R^{(5)}}$, while $g_{\mu\nu}$ is induced metric on the brane with
induced Ricci scalar $R$.\, $\lambda$ is the brane tension (constant
energy density) and $r$ is the crossover scale that is defined as
follows
\begin{equation}
r=\frac{\kappa_{5}^{2}}{2 \kappa_{4}^{2}}=\frac{M_{4}^{2}}{2
M_{5}^{3}}.
\end{equation}
The generalized cosmological dynamics of this setup is given by the
following Friedmann equation [19-21]
\begin{equation}
\varepsilon\sqrt{H^{2}-\frac{
{\cal{M}}}{a^{4}}-\frac{\Lambda_{5}}{6}+\frac{K}{a^{2}}}=r
\alpha(\phi)\Big(H^{2}+\frac{K}{a^{2}}\Big)-
\frac{\kappa_{5}^{2}}{6}(\rho+ \rho_{\phi}+\lambda)\,,
\end{equation}
where $\varepsilon=\pm 1$ are corresponding to two possible branches
of the DGP cosmology and ${\cal{M}}$ is the bulk black hole mass
which is related to the bulk Weyl tensor. The DGP limit has a
Minkowski bulk $\Lambda_{5}=0$ with ${\cal{M}}=0$. We note that
braneworld model with scalar field minimally or non-minimally
coupled to gravity have been studied extensively [23]. The
introduction of non-minimal coupling (NMC) is forced upon us in many
situations of physical and cosmological interests as addressed in
Ref. [24]. Assuming the following line element
\begin{equation}
ds^{2}=q_{\mu\nu}dx^{\mu}dx^{\nu}+b^{2}(y,t)dy^{2}=-n^{2}(y,t)dt^{2}+
a^{2}(y,t)\gamma_{ij}dx^{i}dx^{j}+b^{2}(y,t)dy^{2},
\end{equation}
where $\gamma_{ij}$ is a maximally symmetric 3-dimensional metric
defined as $\gamma_{ij}=\delta_{ij}+k\frac{x_{i}x_{j}}{1-kr^{2}}$,
the energy density of the non-minimally coupled scalar field on the
brane is given as follows [21,23]
\begin{equation}
\rho_{\phi}=\left[\frac{1}{2}\dot{\phi}^{2}+n^{2}V(\phi)-6\alpha'H\dot{\phi}\right]_{y=0},
\end{equation}
where \,$H=\frac{\dot{a}}{a}$\, is Hubble parameter on the brane,\,
$\alpha'=\frac{d\alpha}{d\phi}$ and $\dot{\phi}=\frac{d\phi}{dt}$.

If we consider a flat ($k=0$) brane with $\lambda=0$ and also a
Minkowski bulk ($\Lambda_{5}=0$, ${\cal{M}}=0$), then we can write
equation (3) as follows
\begin{equation}
H^{2}=\pm \frac{H}{r
\alpha(\phi)}+\frac{\kappa_{4}^{2}}{3\alpha(\phi)}\Big(\rho
+\rho_{0\phi}-6\alpha'H\dot{\phi}\Big),
\end{equation}
where
$\rho_{0\phi}=\left[\frac{1}{2}\dot{\phi}^{2}+n^{2}V(\phi)\right]_{y=0}$.
The DGP model has two branches, i.e \, $\varepsilon=\pm 1$
corresponding to two different embedding of the brane in the bulk.
The behavior of two branches at high energies and low energies are
summarized as follows: \\
In the high energy limit we find
\begin{equation}
\hspace{1.5
cm}DGP(\pm):\,\,\,\,\,\,\,\,H^{2}=\frac{\kappa_{4}^{2}}{3\alpha(\phi)}(\rho
+\rho_{\phi}),
\end{equation}
while in the low energy limit we have \\
$$ DGP(+):\,\,\,\,\,\,\,\,H\longrightarrow \frac{1}{r
\alpha(\phi)}-2\frac{\kappa_{4}^{2}\alpha'\dot{\phi}}{\alpha(\phi)}$$
\begin{equation}
DGP(-):\,\,\,\,\,\,\,\,H=0.
\end{equation}
In terms of dimensionless variables introduced in [19,20]
\begin{equation}
h=Hr,\,\,\,\mu=\frac{r\kappa_{5}^{2}}{6}\rho,\,\,\,\,
\bar{\mu}=\frac{r\kappa_{5}^{2}}{6}\rho_{0\phi},\,\,\,\,\sigma=\frac{r\kappa_{5}^{2}}{6}\lambda
,\,\,\,\,\tau=\frac{t}{r},
\end{equation}
we find
\begin{equation}
h^{2}=\pm \frac{h}{
\alpha(\phi)}+\frac{(\mu+\bar{\mu})}{\alpha(\phi)}-\frac{2 h
\kappa_{4}^{2}}{\alpha(\phi)}\frac{d \alpha}{d \tau}.
\end{equation}
The solutions of this equation for $h$ are as follows
\begin{equation}
h=\pm\frac{1}{2\alpha(\phi)}-\frac{\kappa_{4}^{2}}{\alpha(\phi)}\frac{d
\alpha}{d \tau}+\frac{\sqrt{1\mp 4\kappa_{4}^{2}\frac{d \alpha}{d
\tau}+4(\kappa_{4}^{2}\frac{d \alpha}{d
\tau})^{2}+4\alpha(\phi)(\mu+\bar{\mu}+\sigma)}}{2\alpha(\phi)}.
\end{equation}
Here the negative root is not suitable since in the limit of \,
$(\mu +\bar{\mu})\longrightarrow 0$, with this choice of sign one
cannot recover the low energy limit of the model as given by (7).\\

As we have note, the DGP model is an IR modification of the general
relativity. In the UV limit, stringy effects will play important
role. In this viewpoint, to discuss both UV and IR limit of a
cosmologically viable braneworld scenario simultaneously, the DGP
model is not sufficient and we should incorporate stringy effects
via inclusion of the Gauss-Bonnet term.

\section{Incorporation of the Gauss-Bonnet Effect}
The Gauss-Bonnet term with coupling constant $\beta$ is written as
follows
$$L_{GB}=R^{(5)2}-4R_{ab}^{(5)}R^{(5)ab}+R_{abcd}^{(5)}R^{(5)abcd}$$
where $R^{(5)}$ is the curvature scalar of the 5-dimensional bulk
spacetime. These corrections have origin on stringy effects and the
most general action should involve both Gauss-Bonnet and the
Einstein-Hilbert term in 5D theory. The action of the GBIG
(Gauss-Bonnet (GB) term in the bulk and the Induced Gravity (IG)
term on the brane) scenario in the presence of a non-minimally
coupled scalar field on the brane can be written as follows [21]

$$S=\frac{1}{2\kappa_{5}^{2}}\int d^{5}x\sqrt{-g^{(5)}}\Bigg[
R^{(5)}-2\Lambda_{5}+\beta\Big(R^{(5)2}-4R_{ab}^{(5)}R^{(5)ab}+R_{abcd}^{(5)}R^{(5)abcd}\Big)\Bigg]$$
\begin{equation}
+\Bigg[\frac{r}{2\kappa_{5}^{2}}\int
d^{4}x\sqrt{-g}\bigg(\alpha(\phi) R -2\kappa_{4}^{2} g^{\mu\nu}
\nabla_{\mu}\phi\nabla_{\nu}\phi -4\kappa_{4}^{2}V(\phi) -4
\kappa_{4}^{2}\lambda\bigg)\Bigg]_{y=0},
\end{equation}
where $\beta$ and $r$ are the GB coupling constant and IG cross-over
scale respectively. The relation for energy conservation on the
brane is as follows
\begin{equation}
\dot{\rho}+\dot{\rho}_{\phi}+3H(1+\omega)(\rho+\rho_{\phi})=6\alpha'\dot\phi
\Big(H^2+\frac{k}{a^2}\Big).
\end{equation}
where $\omega=\frac{p+p_{\phi}}{\rho+\rho_{\phi}}$\, with $p$ and
$\rho$ pressure and density of the ordinary matter on the brane. The
cosmological dynamics of the model is given by the following
generalized Friedmann equation
\begin{equation}
\Bigg[1+\frac{8}{3}\beta
\Big(H^{2}+\frac{\Psi}{2}+\frac{K}{a^2}\Big)\Bigg]^{2}\Big(H^2-\Psi+\frac{K}{a^2}\Big)=\Bigg[r
\alpha(\phi)H^{2}+r
\alpha(\phi)\frac{K}{a^2}-\frac{\kappa_{5}^{2}}{6}(\rho+\rho_{\phi}+\lambda)
\Bigg]^{2}.
\end{equation}
This equation describes the cosmological evolution on the brane with
tension and a non-minimally coupled scalar field on the brane. The
bulk contains a black hole mass and a cosmological constant. $\Psi$
is defined as follows
\begin{equation}
\Psi+2 \beta \Psi^{2}=\frac{\Lambda_{5}}{6}+\frac{{\cal{M}}}{a^{4}}.
\end{equation}
If $\beta=0$, the model reduces to DGP model, while for $r=0$ we
recover the Gauss-Bonnet model. Here we restrict our study to the
case where bulk black hole mass vanishes, (${\cal{M}}=0$) and
therefore $\Psi+2 \beta \Psi^{2}=\frac{\Lambda_{5}}{6}$. The bulk
cosmological constant in the presence of GB term is given by
$\Lambda_{5}=-\frac{6}{l^{2}}+\frac{12\beta}{l^{4}}$, where $l$ is
the bulk curvature. For a spatially flat brane ($k=0$), the
Friedmann equation is given by
\begin{equation}
\Bigg[1+\frac{8}{3}\beta
\Big(H^{2}+\frac{\Psi}{2}\Big)\Bigg]^{2}\Big(H^2-\Psi\Big)=\Bigg[r
\alpha(\phi)H^{2}-\frac{\kappa_{5}^{2}}{6}(\rho+\rho_{\phi}+\lambda)
\Bigg]^{2}.
\end{equation}
We define the following dimensionless quantities
\begin{equation}
\gamma=\frac{8\beta}{3r^{2}},\hspace{1cm}\chi=\frac{r^{2}}{l^{2}},\hspace{1cm}\psi=\Psi
r^{2},
\end{equation}
where the dimensionless Friedmann equation takes the following form
\begin{equation}
\Bigg[1+\gamma
\Big(h^{2}+\frac{\psi}{2}\Big)\Bigg]^{2}\Big(h^2-\psi\Big)=\Bigg[
\alpha(\phi)h^{2}-\Big(\mu+ \bar{\mu}+\sigma
-2\frac{d\alpha(\phi)}{d\tau}h \kappa_{4}^{2}\Big) \Bigg]^{2}.
\end{equation}
To find cosmological dynamics of our model, we should solve this
equation in an appropriate parameter space. In which follows, we
consider a general case with brane tension ($\sigma\neq0$) and AdS
bulk. Before proceeding further, we should stress on two important
points here: firstly, the presence of GB term removes the big bang
singularity in this setup, and the universe starts with an initial
finite density [19,20]. Gauss-Bonnet effect is essentially a
string-inspired effect in the bulk which its combination with pure
DGP scenario leads to a finite big bang proposal on the brane.
Secondly, non-minimal coupling of the scalar field and induced
gravity on the brane controls the value of the initial density [21].

\subsection{A Brane with Non-Vanishing Tension Embedded in an AdS Bulk}
In the case with $\psi\neq0$, the bulk is AdS since in this case
$\Lambda_{5}\neq 0$. We should solve equation (18) for this case.
The condition $h=0$ gives two solutions
\begin{equation}
\mu_{b,c}=\mp\sqrt{-\psi}(1+\frac{\psi}{2})-\sigma,
\end{equation}
where $\mu_{c}$ is the density at the point that the solution
corresponding to plus sign collapses, while $\mu_{b}$ is the density
of the bouncing point for solution corresponding to minus sign. Here
this point separates into bouncing and collapsing points [21]. To
obtain turning points of the branches, we calculate
$\frac{d(\mu+\bar{\mu})}{d(h^2)}$ as follows
$$\frac{d(\mu+\bar{\mu})}{d(h^2)}=-\frac{\Big[1+\gamma (h^{2}+\frac{\psi}{2})\Big]\Big[3\gamma
(h^{2}-\frac{\psi}{2})+1\Big]}{2\Big(\alpha(\phi)h^{2}-(\mu+\bar{\mu}+\sigma)+2\frac{d\alpha(\phi)}{d\tau}h
\kappa_{4}^{2}\Big)}\hspace{5cm}$$
\begin{equation}
+\frac{2\Bigg[\alpha(\phi)^{2}h^{2}+3\alpha(\phi)\frac{d\alpha(\phi)}{d\tau}h
\kappa_{4}^{2}-(\mu+\bar{\mu}+\sigma)\Big(\alpha(\phi)+\frac{1}{h}\frac{d\alpha(\phi)}{d\tau}
\kappa_{4}^{2}\Big)+2(\frac{d\alpha(\phi)}{d\tau}
\kappa_{4}^{2})^{2}\Bigg]}{2\Big(\alpha(\phi)h^{2}-(\mu+\bar{\mu}+\sigma)+2\frac{d\alpha(\phi)}{d\tau}h
\kappa_{4}^{2}\Big)}.
\end{equation}
By substituting $\frac{d(\mu+\bar{\mu})}{d(h^2)}=0$, these points
can be obtained by solving the equation
\begin{equation} \frac{3}{2}\gamma
h^{3}+(\frac{1}{2}-\frac{3}{4}\gamma \psi) h-\alpha
h\sqrt{h^{2}-\psi}-\frac{d \alpha}{d
\tau}\kappa_{4}^{2}\sqrt{h^{2}-\psi}=0.
\end{equation}
There are three roots, two of which are complex. The real root is
given as follows
$$h=\frac{A}{3(3\gamma-\alpha)}-\Big(6(3\gamma-\alpha)(2-3\gamma\psi+2\alpha\psi)-4A^{2}\Big)\times$$
 $$\Bigg\{3\times 2^{\frac{2}{3}}(3\gamma-\alpha) \bigg(-216\gamma
A-1620 \gamma^{2}\psi A+ 72\alpha A+ 972\gamma\psi\alpha
A-144\psi\alpha^{2} A+16 A^{3}+$$
$$\Big[4\Big(6(3\gamma-\alpha)(2-3\gamma\psi+2\alpha\psi)-4A^{2}\Big)^{3}+(-216\gamma
A-1620 \gamma^{2}\psi A+72\alpha A+972\gamma\psi\alpha
A-144\psi\alpha^{2} A$$
$$+16
A^{3})^{2}\Big]^{\frac{1}{2}}\bigg)^{\frac{1}{3}}\Bigg\}^{-1}+\frac{1}{6\times
2^{\frac{1}{3}}(3\gamma-\alpha)}\Bigg(-216\gamma A-1620
\gamma^{2}\psi A+ 72\alpha A+ 972\gamma\psi\alpha
A-144\psi\alpha^{2} A$$
$$+16 A^{3}+\Big[4\Big(6(3\gamma-\alpha)(2-3\gamma\psi+2\alpha\psi)-4A^{2}\Big)^{3}+(-216\gamma
A-1620 \gamma^{2}\psi A+72\alpha A+972\gamma\psi\alpha A$$
\begin{equation}
-144\psi\alpha^{2}
A+16A^{3})^{2}\Big]^{\frac{1}{2}}\Bigg)^{\frac{1}{3}}\,,
\end{equation}
where $A\equiv\frac{d\alpha}{d\tau}\kappa_{4}^{2}$. Using equation
(18), $h_{\infty}$ ( the re-scaled Hubble parameter on the brane at
late time) for AdS bulk satisfies the following equation
$$h_{\infty}^{6}+\frac{(2\gamma-\alpha^{2})}{\gamma^{2}}h_{\infty}^{4}-\frac{(4\alpha
\frac{d \alpha}{d
\tau}\kappa_{4}^{2})}{\gamma^{2}}h_{\infty}^{3}+\frac{\Big[1+2\alpha\sigma-\psi\gamma(1+\frac{3}{4}\psi\gamma)
-4(\frac{d \alpha}{d
\tau}\kappa_{4}^{2})^{2}\Big]}{\gamma^{2}}h_{\infty}^{2}$$
\begin{equation}
+\frac{(4\sigma \frac{d \alpha}{d
\tau}\kappa_{4}^{2})}{\gamma^{2}}h_{\infty}-\frac{\Big[\psi(1+\frac{\psi\gamma}{2})^{2}+\sigma^{2}\Big]}
{\gamma^{2}}=0.
\end{equation}
This equation has four non-zero roots two of which are negative and
therefore unacceptable on physical ground. When $\gamma\rightarrow
0$, we should recover the non-minimal DGP model ( see the papers by
Nozari in Ref. [23]). From equation (18) one can deduce
\begin{equation}
\mu+\bar{\mu}=\alpha h^{2}-\sigma+2\frac{d \alpha}{d
\tau}h\kappa_{4}^{2}-[1+\gamma(h^{2}+\frac{\psi}{2})](h^{2}-\psi)^{\frac{1}{2}},
\end{equation}
where $h_{\infty}\leq h<h_{i}$ that $h_{i}$ is the re-scaled Hubble
parameter in the beginning ( initial state) of the universe. Since
$(\mu+\bar{\mu})_{\infty}$=0, from equation (24) it follows that
\begin{equation}
\gamma=\frac{\alpha h_{\infty}^{2}-\sigma+2\frac{d \alpha}{d
\tau}\kappa_{4}^{2}h_{\infty}-(h_{\infty}^{2}-\psi)^{\frac{1}{2}}}{(h_{\infty}^{2}+
\frac{\psi}{2})(h_{\infty}^{2}-\psi)^{\frac{1}{2}}}
\end{equation}
By expanding  $(\mu+\bar{\mu})$  to first order in
$h^{2}-h_{\infty}^{2}$, we find
$$\mu+\bar{\mu}=\frac{(h^{2}-h_{\infty}^{2})}{(h_{\infty}^{2}+\frac{\psi}{2})
(h_{\infty}^{2}-\psi)}\Bigg\{(h_{\infty}^{2}+\frac{\psi}{2})(h_{\infty}^{2}-\psi)\big[\alpha+\frac{d
\alpha}{d \tau}\kappa_{4}^{2}h_{\infty}^{-1}\big]$$
$$+(h_{\infty}^{2}-\psi)\big[-\alpha h_{\infty}^{2}+\sigma-2\frac{d
\alpha}{d
\tau}\kappa_{4}^{2}h_{\infty}+(h_{\infty}^{2}-\psi)^{\frac{1}{2}}\big]
+\frac{1}{2}\sigma(h_{\infty}^{2}+\frac{\psi}{2})-(h_{\infty}^{2}+\frac{\psi}{2})\big[\frac{1}{2}\alpha
h_{\infty}^{2}+\frac{d \alpha}{d
\tau}\kappa_{4}^{2}h_{\infty}\big]\Bigg\}.$$ Therefore, the
Friedmann equation can be written as
$$h^{2}=h_{\infty}^{2}+(h_{\infty}^{2}+\frac{\psi}{2})(h_{\infty}^{2}-\psi)
(\mu+\bar{\mu})\Bigg\{(h_{\infty}^{2}+\frac{\psi}{2})(h_{\infty}^{2}-\psi)\big[\alpha+\frac{d
\alpha}{d \tau}\kappa_{4}^{2}h_{\infty}^{-1}\big]$$
$$+(h_{\infty}^{2}-\psi)[-\alpha
h_{\infty}^{2}+\sigma-2\frac{d \alpha}{d
\tau}\kappa_{4}^{2}h_{\infty}+(h_{\infty}^{2}-\psi)^{\frac{1}{2}}]+
\frac{1}{2}\sigma(h_{\infty}^{2}+\frac{\psi}{2})-(h_{\infty}^{2}+\frac{\psi}{2})[\frac{1}{2}\alpha
h_{\infty}^{2}+\frac{d \alpha}{d
\tau}\kappa_{4}^{2}h_{\infty}]\Bigg\}^{-1}.$$ In comparison with
equation (6), we find the following effective 4-dimensional Newton's
constant
$$G_{eff}=(h_{\infty}^{2}+\frac{\psi}{2})(h_{\infty}^{2}-\psi)
\Bigg[2\Big\{(h_{\infty}^{2}+\frac{\psi}{2})(h_{\infty}^{2}-\psi)
[\alpha+\frac{d \alpha}{d \tau}\kappa_{4}^{2}h_{\infty}^{-1}]$$
$$+(h_{\infty}^{2}-\psi)[-\alpha h_{\infty}^{2}+\sigma-2\frac{d
\alpha}{d\tau}\kappa_{4}^{2}h_{\infty}+(h_{\infty}^{2}-\psi)^{\frac{1}{2}}]$$
$$+\frac{1}{2}\sigma(h_{\infty}^{2}+\frac{\psi}{2})-(h_{\infty}^{2}+
\frac{\psi}{2})[\frac{1}{2}\alpha h_{\infty}^{2}+ \frac{d \alpha}{d
\tau}\kappa_{4}^{2}h_{\infty}]\Big\}\Bigg]^{-1}{\alpha G_{5}\over
r}$$ where $G_{5}=\frac{\kappa_{5}^{2}}{8\pi}$ and
$G_{eff}=\frac{\kappa_{4}^{2}}{8\pi}$ are five and four dimensional
gravitational constant respectively so that
$$M_{5}^{3}=(r^{2}H_{0}^{2}+\frac{\psi}{2})(r^{2}H_{0}^{2}-
\psi)\Bigg[2\Big\{(r^{2}H_{0}^{2}+\frac{\psi}{2})(r^{2}H_{0}^{2}-
\psi)[\alpha+\frac{d \alpha}{d
\tau}\kappa_{4}^{2}r^{-1}H_{0}^{-1}]$$
$$+(r^{2}H_{0}^{2}-\psi)[-\alpha r^{2}H_{0}^{2}+\sigma-2\frac{d
\alpha}{d\tau}\kappa_{4}^{2}rH_{0}+(r^{2}H_{0}^{2}-\psi)^{\frac{1}{2}}]$$
$$+\frac{1}{2}\sigma(r^{2}H_{0}^{2}+\frac{\psi}{2})-(r^{2}H_{0}^{2}+
\frac{\psi}{2})[\frac{1}{2}\alpha r^{2}H_{0}^{2}+ \frac{d \alpha}{d
\tau}\kappa_{4}^{2}rH_{0}]\Big\}\Bigg]^{-1}{\alpha M_{p}^{2}\over
r}.$$ We note that the solution with minus sign in equation (19)
gives a bouncing cosmological solution. A bouncing universe goes
from an era of accelerated collapse to an expanding phase without
displaying any singularity. In a bouncing universe, during the
contracting phase, the scale factor a(t) is decreasing, i.e.
$\dot{a}(t) < 0$, and in the expanding phase we have  $\dot{a}(t) >
0$. At the bouncing point, $\dot{a}(t) = 0$, and around this point
$\ddot{a}(t)
> 0$ for a period of time. In the bouncing universe, the equation of state
parameter of the matter content, $\omega$, must transit from
$\omega<-1$ to $\omega>-1$ [25] ( see also [26]). In our framework,
as we will see, in a suitable domain of the parameters space, by
increasing $\alpha(\phi)$ values the solution with minus sign in the
equation (19) contains a bouncing cosmology with a transition from
$\omega<-1$ to $\omega>-1$. However this solution disappear for some
values of $\psi$. For instance, our inspection shows that with
$\psi=-5$, this bouncing solution disappears completely. These
arguments show that inclusion of the non-minimal coupling of the
scalar field and induced gravity on the brane can be used to
fine-tune this braneworld scenario to realize bouncing solution in
appropriate domain of the parameters space of the model.

\section{A GBIG-Dark Energy Model }
To study a dark energy model in GBIG scenario with a non-minimally
coupled scalar field on the brane, we first present a brief overview
of the holographic dark energy model [27,28]. It is well-known that
the mass of a spherical and uncharged D-dimensional black hole is
related to its Schwarzschild radius by [28]
\begin{equation}
M_{BH}=r_{s}^{D-3}\Big(\sqrt{\pi}M_{D}\Big)^{D-3}M_{D}
\frac{D-2}{8\Gamma\Big(\frac{D-1}{2}\Big)}
\end{equation}
where the D-dimensional Planck mass, $M_{D}$, is related to the
D-dimensional gravitational constant $G_{D}$ and the usual
4-dimensional Planck mass through $ M_{D}=G_{D}^{-\frac{1}{D-2}}$
and $M_{p}^{2}=M_{D}^{D-2}V_{D-4}$ with $V_{D-4}$ the volume of the
extra-dimensional space. If $\rho_{\Lambda D}$ is the bulk vacuum
energy, then application of the holographic dark energy proposal in
the bulk gives
\begin{equation}
\rho_{\Lambda D} Vol(S^{D-2})\leq
r_{D-3}\Big(\sqrt{\pi}M_{D}\Big)^{D-3}M_{D}
\frac{D-2}{8\Gamma\Big(\frac{D-1}{2}\Big)}
\end{equation}
where $Vol(S^{D-2})$ is the volume of the maximal hypersphere in a
D-dimensional spacetime given by $Vol(S^{D-2})=A_{D}r^{D-1}$.
$A_{D}$ is defined as
$$A_{D}=\frac{\pi^{\frac{D-1}{2}}}{\Big(\frac{D-1}{2}\Big)!}$$
$$A_{D}=\frac{\Big(\frac{D-2}{2}\Big)!}{(D-1)!}2^{D-1}\pi^{\frac{D-2}{2}}$$
for $D-1$ being even or odd respectively. By saturating inequality
(27), introducing $L$ as a suitable large distance ( IR cutoff) and
$c^{2}$ as a numerical factor, the corresponding vacuum energy as a
holographic dark energy is given by [28]
\begin{equation}
\rho_{\Lambda D}=c^{2}\Big(\sqrt{\pi}M_{D}\Big)^{D-3}M_{D}A_{D}^{-1}
\frac{D-2}{8\Gamma\Big(\frac{D-1}{2}\Big)}L^{-2}.
\end{equation}
Using this expression, one can calculate the corresponding pressure
via continuity equation and then the equation of state parameter of
the holographic dark energy defined as
$\omega_{\Lambda}=\frac{\rho_{\Lambda}}{p_{\Lambda}}$ can be
obtained directly.

Now we use this formalism in our GBIG setup. We note that in our
forthcoming argument we choose the IR cut-off,\, $L$ \,, to be the
crossover scale $r$ which is related to the Hubble radius ( see [28]
for alternative possibilities). In this case the holographic dark
energy density is given by
\begin{equation}
\rho_{\phi}=3(8\pi G_{eff})^{-1}\frac{h^{2}}{r^{2}}.
\end{equation}
Using the value of \, $G_{eff}$\,  obtained previously, the energy
density of the dark energy component on the brane is given by
$$\rho_{\phi}=\frac{(\frac{6h^{2}}{\kappa_{5}^{2}r\alpha})}
{(h_{\infty}^{2}+\frac{\psi}{2})(h_{\infty}^{2}-\psi)}
\Bigg\{(h_{\infty}^{2}+\frac{\psi}{2})(h_{\infty}^{2}-
\psi)[\alpha+\frac{d \alpha}{d \tau}\kappa_{4}^{2}h_{\infty}^{-1}]$$
$$+ (h_{\infty}^{2}-\psi)[-\alpha h_{\infty}^{2}+ \sigma-2\frac{d
\alpha}{d \tau}\kappa_{4}^{2}h_{\infty}+
(h_{\infty}^{2}-\psi)^{\frac{1}{2}}]$$
$$+\frac{1}{2}\sigma(h_{\infty}^{2}+
\frac{\psi}{2})-(h_{\infty}^{2}+\frac{\psi}{2})[\frac{1}{2}\alpha
h_{\infty}^{2}+ \frac{d \alpha}{d
\tau}\kappa_{4}^{2}h_{\infty}]\Bigg\},$$ and the corresponding
pressure can be calculated as follows
$$p_{\phi}=-\frac{2}{\kappa_{5}^{2}\alpha}(h_{\infty}^{2}+
\frac{\psi}{2})^{-1}(h_{\infty}^{2}-\psi)^{-1}\Bigg[\big(\frac{2\alpha\dot{h}-
\frac{d \alpha}{d
\tau}h}{\alpha}+\frac{3h^{2}}{r}\big)\Big\{(h_{\infty}^{2}+
\frac{\psi}{2})(h_{\infty}^{2}-\psi)\big[\alpha+\frac{d \alpha}{d
\tau}\kappa_{4}^{2}h_{\infty}^{-1}\big]$$
$$(h_{\infty}^{2}-\psi)\big[-\alpha h_{\infty}^{2}+\sigma-2\frac{d \alpha}{d \tau}\kappa_{2}^{4}h_{\infty}+
(h_{\infty}^{2}-\psi)^{\frac{1}{2}}\big]+\frac{1}{2}\sigma(h_{\infty}^{2}+\frac{\psi}{2})-(h_{\infty}^{2}+
\frac{\psi}{2})\big[\frac{1}{2}\alpha h_{\infty}^{2}+\frac{d
\alpha}{d \tau}\kappa_{2}^{4}h_{\infty}\big]\Big\}$$
$$+h\Big\{(h_{\infty}^{2}+\frac{\psi}{2})(h_{\infty}^{2}-\psi)\big[\frac{d \alpha}{d \tau}+
\frac{d^{2} \alpha}{d \tau^{2}}\kappa_{4}^{2}h_{\infty}^{-1}\big]+
(h_{\infty}^{2}-\psi)\big[-\frac{d \alpha}{d
\tau}h_{\infty}^{2}-2\frac{d^{2} \alpha} {d
\tau^{2}}\kappa_{4}^{2}h_{\infty}\big]$$
$$-(h_{\infty}^{2}+\frac{\psi}{2})\big[\frac{1}{2}\frac{d \alpha}{d \tau}h_{\infty}^{2}+
\frac{d^{2} \alpha}{d
\tau^{2}}\kappa_{4}^{2}h_{\infty}\big]\Big\}\Bigg]+2\frac{\alpha'\dot{\phi}h}{r}.$$
Now the equation of state parameter of the model is given by
$$\omega_{\Lambda}=-\frac{r}{3h^{2}}\frac{2\alpha\dot{h}-
\frac{d \alpha}{d
\tau}}{\alpha}-1-\frac{r}{3h}\Bigg\{\bigg((h_{\infty}^{2}+
\frac{\psi}{2})(h_{\infty}^{2}-\psi)\big[\frac{d \alpha}{d \tau}+
\frac{d^{2} \alpha}{d \tau^{2}}\kappa_{4}^{2}h_{\infty}^{-1}\big]+
(h_{\infty}^{2}-\psi)\big[-\frac{d \alpha}{d \tau}h_{\infty}^{2}-
2\frac{d^{2} \alpha}{d \tau^{2}}\kappa_{4}^{2}h_{\infty}\big]$$
$$-(h_{\infty}^{2}+\frac{\psi}{2})\big[\frac{1}{2}\frac{d \alpha}
{d \tau}h_{\infty}^{2}+\frac{d^{2} \alpha}{d
\tau^{2}}\kappa_{4}^{2}h_{\infty}\big]
\bigg)\bigg((h_{\infty}^{2}+\frac{\psi}{2})(h_{\infty}^{2}-\psi)\big[\alpha+
\frac{d \alpha}{d
\tau}\kappa_{4}^{2}h_{\infty}^{-1}\big]+(h_{\infty}^{2}-
\psi)\big[-\alpha h_{\infty}^{2}+\sigma-2\frac{d \alpha}{d
\tau}\kappa_{4}^{2}h_{\infty}$$
$$+(h_{\infty}^{2}-\psi)^{\frac{1}{2}}\big]+\frac{1}{2}\sigma(h_{\infty}^{2}+
\frac{\psi}{2})-(h_{\infty}^{2}+\frac{\psi}{2})\big[\frac{1}{2}\alpha
h_{\infty}^{2}+ \frac{d \alpha}{d
\tau}\kappa_{4}^{2}h_{\infty}\big]\bigg)^{-1}\Bigg\}
+\frac{\alpha'\dot{\phi}\alpha\kappa_{5}^{2}}{3h}(h_{\infty}^{2}+\frac{\psi}{2})(h_{\infty}^{2}-\psi)$$
$$\Bigg\{(h_{\infty}^{2}+\frac{\psi}{2})(h_{\infty}^{2}-
\psi)[\alpha+\frac{d \alpha}{d \tau}\kappa_{4}^{2}h_{\infty}^{-1}]+
(h_{\infty}^{2}-\psi)[-\alpha h_{\infty}^{2}+ \sigma-2\frac{d
\alpha}{d \tau}\kappa_{4}^{2}h_{\infty}+
(h_{\infty}^{2}-\psi)^{\frac{1}{2}}]$$
$$+\frac{1}{2}\sigma(h_{\infty}^{2}+
\frac{\psi}{2})-(h_{\infty}^{2}+\frac{\psi}{2})[\frac{1}{2}\alpha
h_{\infty}^{2}+ \frac{d \alpha}{d
\tau}\kappa_{4}^{2}h_{\infty}]\Bigg\}^{-1}.$$ This is a complicated
equation that cannot be explained easily. To proceed further, we set
$ a(t)=a_{0} t^{\nu}$ and $\phi(t)=\phi_{0} t^{-\mu}$. This is a
reliable ansatz which is physically well-motivated since it is
corresponding to an accelerating universe if $\nu>1$ and decreasing
scalar field for positive $\mu$. Choosing appropriate values of the
parameters of the model ( for instance we have set $\nu=1.2$\,,
$\mu=0.5$\, , $\xi=0.15$,\, $r=1.26 H_{0}^{-1}\equiv 1$\,,
$k_{4}^{2}$=1\,, $k_{5}^{2}=1$\,, $\sigma=10$ and $\psi=-1$),
dynamics of the equation of state parameter of the model is shown by
the figure $1$. As this figure shows, equation of state parameter of
the model crosses the phantom divide line from less than $-1$ to its
above in the favor of bouncing solution. We note that due to wide
parameter space of this model, it is possible to realize phantom
divide line crossing from above $-1$ to its below more acceptable on
the basis of the recent observations. This possibility is shown in
figure $2$.
\begin{figure}[htp]
\begin{center}\includegraphics{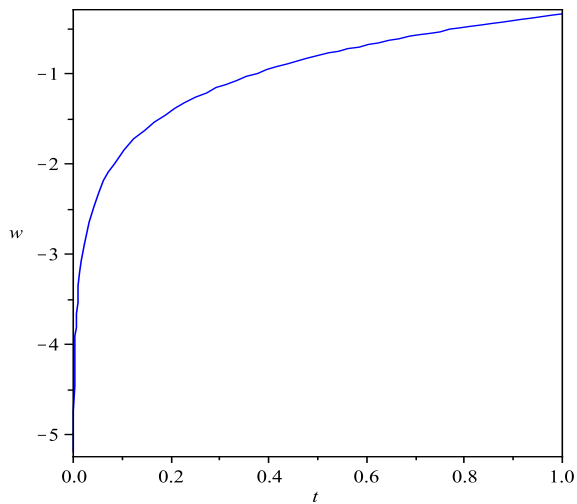} \vspace{7cm}
\end{center}
 \caption{\small {Dynamics of the equation of state parameter.
 It is possible to realize phantom divide line crossing in this setup.
 The crossing is possible from $\omega<-1$ to $\omega>-1$,
 a typical behavior of bouncing solutions.}}
\end{figure}

\begin{figure}[htp]
\begin{center}\includegraphics{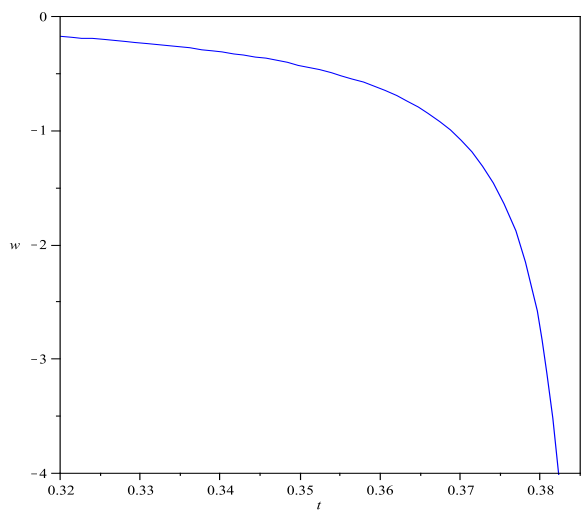} \vspace{7cm}
\end{center}
 \caption{\small {Crossing of the phantom divide line from
 quintessence ($\omega>-1$) to phantom phase ($\omega<-1$).
 This type of crossing behavior is supported observationally. }}
\end{figure}

\newpage
\section{ The Minimal Theory}
To have a more complete discussion, we study dynamics of the
equation of state parameter for a GBIG theory where a quintessence
field is minimally coupled to induced gravity on the brane.

\subsection{Minkowski Bulk with a Tensionless Brane}

For the case with Minkowski bulk ($\psi=0$) and a tensionless brane
($\sigma=0$), the Friedmann equation is given as follows [20,21]
\begin{equation}
(1+\gamma h^{2})^{2}h^{2}=\bigg[h^{2}-(\mu+\bar{\mu})\bigg]^{2},
\end{equation}
from which we can deduce
\begin{equation}
\mu+\bar{\mu}=h^{2}-h(\gamma h^{2}+1).
\end{equation}
Since\ $(\mu+\bar{\mu})_{\infty}=0$,\ from equation (31) it follows
that
\begin{equation}
\gamma=\frac{h_{\infty}-1}{h_{\infty}^{2}}.
\end{equation}
We note that $\gamma$ is constraint by recent observational data
from SNIa+LSS+H(z) combined dataset so that
$\gamma=0.000_{-0.000}^{+0.003}$\,\, [29]. By expanding\
$\mu+\bar{\mu}$\ to first order in \, $h^{2}-h_{\infty}^{2}$, we
find
\begin{equation}
h^{2}=h_{\infty}^{2}+\frac{2h_{\infty}}{2-h_{\infty}}(\mu+\bar{\mu}).
\end{equation}
Then we can obtain the following effective Newton's constant
\begin{equation}
G_{eff}=\Big(\frac{h_{\infty}}{2-h_{\infty}}\Big)\frac{G_{5}}{r},
\end{equation}
where $G_{5}=\kappa_{5}^{2}/8\pi$\ and
$G_{eff}=\kappa_{4}^{2}/8\pi$\ are five-and four-dimensional
gravitational constants respectively. From this equation we obtain
\begin{equation}
M_{5}^{3}\simeq\Big(\frac{h_{\infty}}{2-h_{\infty}}\Big)\frac{M_{p}^{2}}{r}.
\end{equation}
The effective dark energy density is given by
\begin{equation}
\rho=3(8\pi G_{eff})^{-1}\frac{h^{2}}{r^{2}}.
\end{equation}
So from equation (34) we obtain
\begin{equation}
\rho=\frac{3}{r\kappa_{5}^{2}}(\frac{2-h_{\infty}}{h_{\infty}})
\end{equation}
Using the conservation equation, $\dot{\rho}+3H\rho(1+\omega)=0$,\,
we find
\begin{equation}
p=-\frac{1}{\kappa_{5}^{2}}(2\dot{h}+\frac{3h^{2}}{r})(\frac{2-h_{\infty}}{h_{\infty}}).
\end{equation}
Hence, the equation of state parameter of the model is given as
follows
\begin{equation}
\omega=-1-\frac{2r\dot{h}}{3h^{2}}
\end{equation}
This equation of state parameter has the potential to describe a
variety of cosmologically interesting cases. For instance, for a
model universe with $a(t)\propto \ln t$ where $t$ is the cosmic
time, this model has the potential to realize phantom divide line
crossing. Also for models such as
$a(t)=a_{0}\Big(t^{2}+\frac{t_{0}}{1-\nu}\Big)^{\frac{1}{1-\nu}}$,
that are phenomenologically reliable [30], this crossing is possible
too ( see figure $2$).

\subsection{AdS bulk with brane tension}
For the case with AdS bulk ($\psi\neq 0$) and brane tension
($\sigma\neq0$), the Friedmann equation is as follows
\begin{equation}
\bigg[1+\gamma(h^{2}+\frac{\psi}{2})\bigg]^{2}(h^{2}-\psi)=\bigg[h^{2}-(\mu+\bar{\mu}+\sigma)\bigg]^{2},
\end{equation}
which can be rewritten as follows
\begin{equation}
\mu+\bar{\mu}=h^{2}-\sigma-\bigg[1+\gamma(h^{2}+\frac{\psi}{2})\bigg]
(h^{2}-\psi)^{\frac{1}{2}}.
\end{equation}
Since\ $(\mu+\bar{\mu})_{\infty}=0$\,, from this equation we find
\begin{equation}
\gamma=\frac{h_{\infty}^{2}-\sigma-(h_{\infty}^{2}-\psi)^{\frac{1}{2}}}{(h_{\infty}^{2}+\frac{\psi}{2})
(h_{\infty}^{2}-\psi)^{\frac{1}{2}}}.
\end{equation}
By expanding \ $\mu+\bar{\mu}$\ to first order in \
$h^{2}-h_{\infty}^{2}$,\ we find
\begin{equation}
\mu+\bar{\mu}=(h^{2}-h_{\infty}^{2})(h_{\infty}^{2}+\frac{\psi}{2})^{-1}(h_{\infty}^{2}-\psi)^{-1}
\Bigg\{\frac{1}{2}(h_{\infty}^{2}+\frac{\psi}{2})[\sigma+h_{\infty}^{2}-2\psi]+(h_{\infty}^{2}-\psi)
[-h_{\infty}^{2}+\sigma+(h_{\infty}^{2}-\psi)^{\frac{1}{2}}]\Bigg\},
\end{equation}
which can be rewritten as follows
\begin{equation}
h^{2}=h_{\infty}^{2}+(h_{\infty}^{2}+\frac{\psi}{2})(h_{\infty}^{2}-\psi)(\mu+\bar{\mu})\Bigg\{\frac{1}{2}
(h_{\infty}^{2}+\frac{\psi}{2})(\sigma+h_{\infty}^{2}-2\psi)(h_{\infty}^{2}-\psi)
[-h_{\infty}^{2}+\sigma+(h_{\infty}^{2}-\psi)^{\frac{1}{2}}]\Bigg\}^{-1}.
\end{equation}
By comparing this relation with equation (6), we find the following
relation for 4-dimensional effective Newtonian gravitational
constant
\begin{equation}G_{eff}=\Bigg\{\frac{1}{2}(h_{\infty}^{2}+\frac{\psi}{2})(h_{\infty}^{2}-\psi)
\Bigg(\frac{1}{2}(h_{\infty}^{2}+\frac{\psi}{2})(\sigma+h_{\infty}^{2}-2\psi)(h_{\infty}^{2}-\psi)
[-h_{\infty}^{2}+\sigma+(h_{\infty}^{2}-\psi)^{\frac{1}{2}}]\Bigg)^{-1}\Bigg\}\frac{G_{5}}{r}.
\end{equation}
Now the relation between \, $M_{5}^{3}$\, and \,$M_{p}^{2}$ \,
reads as follows
\begin{equation}
M_{5}^{3}=\Bigg(\frac{1}{2}(h_{\infty}^{2}+\frac{\psi}{2})(h_{\infty}^{2}-\psi)
\Bigg\{\frac{1}{2}(h_{\infty}^{2}+\frac{\psi}{2})(\sigma+h_{\infty}^{2}-2\psi)(h_{\infty}^{2}-\psi)
[-h_{\infty}^{2}+\sigma+(h_{\infty}^{2}-\psi)^{\frac{1}{2}}]\Bigg\}^{-1}\Bigg)\frac{M_{p}^{2}}{r}
\end{equation}
The holographic dark energy density and pressure of this model are
given by
\begin{equation}
\rho=\frac{6h^{2}}{\kappa_{5}^{2}r}(h_{\infty}^{2}+\frac{\psi}{2})^{-1}(h_{\infty}^{2}-\psi)^{-1}\Bigg\{\frac{1}{2}
(h_{\infty}^{2}+\frac{\psi}{2})(\sigma+h_{\infty}^{2}-2\psi)(h_{\infty}^{2}-\psi)
[-h_{\infty}^{2}+\sigma+(h_{\infty}^{2}-\psi)^{\frac{1}{2}}]\Bigg\},
\end{equation}
and
\begin{equation}
p=-\frac{2}{\kappa_{5}^{2}}(h_{\infty}^{2}+\frac{\psi}{2})^{-1}(h_{\infty}^{2}-\psi)^{-1}
\Bigg[(2\dot{h}+\frac{3h^{2}}{r})\bigg\{\frac{1}{2}
(h_{\infty}^{2}+\frac{\psi}{2})(\sigma+h_{\infty}^{2}-2\psi)(h_{\infty}^{2}-\psi)
[-h_{\infty}^{2}+\sigma+(h_{\infty}^{2}-\psi)^{\frac{1}{2}}]\bigg\}\Bigg]
\end{equation}
respectively. The equation of state parameter of the holographic
dark energy component now is given by
\begin{equation}
\omega=-1-\frac{2r\dot{h}}{3h^{2}}
\end{equation}
where $h$ is given by equation (44). This model accounts for
crossing of the cosmological constant line if we set for instant
$$a(t)=a_{0}\Big(t^{2}+\frac{t_{0}}{1-\nu}\Big)^{\frac{1}{1-\nu}}.$$
Figure $2$ shows the realization of the phantom divide line crossing
again from $\omega<-1$ to $\omega>-1$ in this setup.
\begin{figure}[htp]
\begin{center}\includegraphics{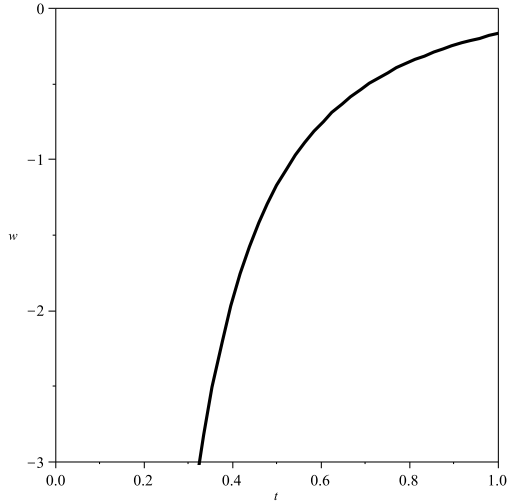} \vspace{6.5cm}
\end{center}
 \caption{\small {Dynamics of the equation of state parameter.
 It is possible to realize phantom divide line crossing in this setup.}}
\end{figure}

We note that one important point in this DGP-inspired model is the
issue of stability of the self-accelerated solutions. It has been
shown that the self-accelerating branch of the DGP model contains a
ghost at the linearized level [31]. The ghost carries negative
energy density and it leads to the instability of the spacetime. The
presence of the ghost can be related to the infinite volume of the
extra-dimension in DGP setup. It is shown also that the introduction
of the Gauss-Bonnet term in the bulk does not help to overcome this
problem [32]. In fact, it is still unclear what is the end state of
the ghost instability in self-accelerated branch of the DGP inspired
models (for more details see [31]). On the other hand, non-minimal
coupling of the scalar field and induced gravity provides a new
degree of freedom which requires special fine tuning and this may
provide a suitable basis to treat ghost instability. Lorentz
invariance violating scenarios are other alternative paradigm which
may shed light in this problem due to their wider parameter spaces
[33]. As we have shown here, non-minimal coupling of the scalar
field and induced gravity has the capability to account for bouncing
solutions. It seems that this additional degree of freedom has also
the capability to provide the background for a reliable solution to
ghost instability. However, it is not trivial at this level and need
further justification.

\section{Summary and Conclusions}
DGP model provides an IR modification of the general relativity. On
the other hand, at the UV limit, stringy effects play important
role. In this viewpoint, to discuss both UV and IR limit of a
braneworld scenario simultaneously, the DGP model is not sufficient
alone and one should incorporate stringy effects via inclusion of
the Gauss-Bonnet term in the bulk action. The presence of GB term
removes the big bang singularity, and the universe starts with an
initial finite density [20,21]. Non-minimal coupling of scalar field
and induced gravity on the brane which is motivated from several
compelling reasons, controls the value of the initial density in a
finite big bang cosmology on the brane [21]. In this paper, we have
constructed a holographic picture of dark energy model by
incorporating Gauss-Bonnet effect in a DGP braneworld model. By
considering the IR cut-off of the model to be the crossover scale of
the theory, we have obtained dynamics of the equation of state
parameter of the holographic dark energy component. By adopting a
suitable ansatz, we have shown that this model accounts for phantom
divide line crossing in a wide range of the parameter space. Then we
have studied the minimal case with detailed by considering two
possible geometry of the bulk manifold and brane tension. In these
case also there is crossing of the phantom divide line by adopting
suitable ansatz. One of the main outcome of our analysis is the
implication of the non-minimal coupling on the bouncing cosmologies.
In the bouncing universe, the equation of state parameter of the
matter content, $\omega$, must transit from $\omega<-1$ to
$\omega>-1$. In our framework, as figure $1$ and $2$ show, this
condition is possible to be realized and therefore our model
essentially supports bouncing solutions. Another important point is
the fact that inclusion of the non-minimal coupling of the scalar
field and induced gravity on the brane in the presence of the
Gauss-Bonnet term could be used to fine-tune braneworld cosmological
models in the favor of the recent observational data. In this manner
it is possible to find a severe constraint on the value of the
non-minimal coupling in the spirit of scalar-tensor theories.
Finally, the issue of ghost instabilities in the self-accelerated
solutions and possible impacts of the Gauss-Bonnet term and
non-minimal coupling on this issue are addressed here.

\end{document}